\documentclass[a4paper, 12pt]{article}
\usepackage{epsfig,amssymb,euscript,xspace, color}
\usepackage{amsmath,jheppub,mathtools,empheq,amsthm}
\usepackage{graphicx}
\usepackage{verbatim}
\usepackage{cancel}

\usepackage[T1]{fontenc} 
\usepackage{tikz,caption,subcaption,marvosym} 
\usetikzlibrary{decorations.markings,arrows,snakes}
\usepackage{epsfig,amssymb,euscript,xspace, color}
\usepackage{amsmath,mathtools,rotating}

\def\bea{\begin{eqnarray}}
\def\eea{\end{eqnarray}}
\def\be{\begin{equation}}
\def\ee{\end{equation}}



\begin{document}

\title{
{\large All chiral ${\cal W}$-algebra extensions of $\mathfrak{so}(2,3)$}}
\author{Nishant Gupta and Nemani V. Suryanarayana}
\affiliation{Institute of Mathematical Sciences,\\ Taramani, Chennai 600113, India \\
\& \\ Homi Bhabha National Institute,  \\ Anushakti Nagar, Mumbai 400094, India }
\emailAdd{nishantg, nemani@imsc.res.in}

\abstract{
We show that there are four chiral ${\cal W}$-algebra extensions of $\mathfrak{so}(2,3)$ algebra and construct them explicitly. We do this by a simple identification of each of the inequivalent embeddings of a copy of $\mathfrak{sl}(2,{\mathbb R})$ in the $\mathfrak{so}(2,3)$ algebra and the maximal subalgebra $\mathfrak{h}$ that commutes with it. 
Then using the standard 2d chiral CFT techniques we find the corresponding ${\cal W}$-algebra extensions. Two of the four resultant ${\cal W}$-algebras are new, one of which may be thought of as the conformal $\mathfrak{bms}_3$ algebra valid for finite values of its central charge.}
\maketitle
\section{Introduction}
Symmetries play a crucial role in our understanding of quantum field theories. More the symmetries, more the control over the physical quantities of that theory. In the context of 2$d$ conformal field theories, it has been well known that the symmetry algebras are not only infinite dimensional but can also be non-linear, leading to the study of ${\cal W}$ algebras in physics (starting from \cite{Zamolodchikov:1985wn}; see, for instance, the reviews \cite{Bouwknegt:1992wg, Blumenhagen:2009zz}). In the context of higher (than two) dimensional conformal field theories, it is not generically expected that the symmetry algebras are infinite dimensional. This general expectation is corroborated by the fact that, at least for a CFT$_d$ with holographic duals, the asymptotic symmetry algebra with Dirichlet boundary conditions proposed by Brown-Henneaux \cite{Brown:1986nw} for $AdS_{d+1}$ gravity is $\mathfrak{so}(2,d)$ for any  $d \ge 3$.\footnote{ The non-Dirichlet boundary conditions that lead to  infinite dimensional extensions of $\mathfrak{so}(2,d)$ as the symmetry algebra of AdS$_{d+1}$ for $d \geq 3$ were considered in \cite{Compere:2019bua, Compere:2020lrt, Fiorucci:2020xto, Gupta:2022mdt}.} However, for the corresponding ${\mathbb R}^{1,d}$ at null-infinities the asymptotic symmetry algebras are expected to be infinite dimensional -- either the $\mathfrak{bms}_{d+1}$ algebra \cite{Bondi:1962px, Sachs:1962zza, Sachs:1962wk} or its various extended versions \cite{Barnich:2009se, Barnich:2010eb, Barnich:2011mi, Campiglia:2014yka, Campiglia:2015yka}. These in turn are interpreted to be the symmetries of the so-called (Euclidean $(d-1)$-dimensional) celestial CFTs. In particular, for $d=3$ it has emerged \cite{Banerjee:2020zlg, Banerjee:2021dlm} that the symmetry algebra relevant is chiral $\mathfrak{bms}_4$ algebra which can be thought of as a chiral conformal algebra of a 2$d$ CFT. This algebra is a chiral extension of $\mathfrak{iso}(1,3)$ and consists of the following six chiral operators: one stress tensor $T(z)$, a triplet of $\mathfrak{sl}(2, {\mathbb R})$ currents $J_a(z)$ ($a=0,\pm 1$) and a doublet $G_s(z)$ ($s=\pm 1/2$) that are conformal ( $h=3/2$) as well as current algebra primaries (transforming in the doublet representation of $\mathfrak{sl}(2, {\mathbb R})$).\footnote{This algebra can be realised as the asymptotic symmetry algebra of 4$d$ Einstein gravity with an appropriate set of asymptotically flat boundary conditions \cite{Gupta:2021cwo}.} In \cite{Gupta:2022mdt} we derived the most general chiral conformal algebra,  chiral $\Lambda$-$\mathfrak{bms}_4$ with this spectrum of operators from the AdS$_4$ gravity in semi-classical limit (large current level $\kappa$). There we also derived the complete quantum version of this algebra using the property of associativity of operator product algebra of $2d$ chiral CFT. The chiral $\Lambda$-$\mathfrak{bms}_4$\footnote{The $\Lambda$-$\mathfrak{bms}_4$ terminology was introduced in the works  \cite{Compere:2019bua, Compere:2020lrt,Fiorucci:2020xto}.} is a non-linear $\mathcal W$-algebra which is a chiral extension of $\mathfrak{so}(2,3)$ algebra and is isomorphic to the $N=1$ case of quasi-superconformal algebras which includes spin-1 Kac-Moody current $\mathfrak{sp}(2)$ derived by Romans in a different context \cite{Romans:1990ta}. This chiral $\Lambda$-$\mathfrak{bms}_4$ algebra admits a contraction to the chiral $\mathfrak{bms}_4$ algebra of \cite{Banerjee:2020zlg, Banerjee:2021dlm}.

In this paper, we ask whether there are other chiral $\mathcal W$-algebra extensions of $\mathfrak{so}(2,3)$ in addition to chiral $\Lambda$-$\mathfrak{bms}_4$ algebra.  It is important to address this question as there are several works in the literature where more than one chiral extension of a finite dimensional algebra played a role in the study of holography. For example in \cite{Avery:2013dja}, the authors derived the infinite dimensional chiral extension of $\mathfrak{so}(2,2)$ containing $\mathfrak{sl}(2,\mathbb R)$ current algebra from the asymptotic symmetries of AdS$_3$ gravity. As we will show in the Appendix \ref{appendixA},  this is one of the only two chiral extensions of $\mathfrak{so}(2,2)$, the other one being isomorphic to Brown-Henneaux algebra \cite{Brown:1986nw}.  Similarly in the context of holography of higher spin gravity in AdS$_3$, it was shown that for spin-3 gravity in $\mathfrak{sl}(3,\mathbb R) \times \mathfrak{sl}(3,\mathbb R)$ Chern-Simons theory formulation, both chiral $\mathcal W$- algebra extensions of $\mathfrak{sl}(3,\mathbb R)$, namely $ W_3$ and Polyakov-Bershadsky algebra $W_3^{(2)}$ gave rise to different types of AdS$_3$ vacua \cite{Ammon:2011nk}. Therefore, it is expected that various chiral $\mathcal W$-algebra extensions of $\mathfrak{so}(2,3)$ will also play a role in AdS$_4$/CFT$_3$ correspondence (like chiral $\Lambda$-$\mathfrak{bms}_4$ algebra) and study of asymptotic symmetries of 3d conformal gravity. 

In this work, we show that the chiral $\Lambda$-$\mathfrak{bms}_4$ algebra found in \cite{Romans:1990ta, Gupta:2022mdt} is one of four such $\mathcal W$-algebras. We identify the remaining three explicitly completing the list. The construction of these $\mathcal W$-algebras follows from these simple steps:  
\begin{enumerate}
	\item To start with, we show that there are exactly four inequivalent ways to embed a copy of $\mathfrak{sl}(2, \mathbb{R})$ inside $\mathfrak{so}(2, 3)$.
	\item Then we identify in each case the maximal sub-algebra $\mathfrak{h} \subset \mathfrak{so}(2, 3)$ that commutes with the
	$\mathfrak{sl}(2, \mathbb{R})$ of the first step.
	\item The rest of the generators arrange themselves into finite dimensional irreducible representations of the sub-algebra $\mathfrak{sl}(2, \mathbb{R}) \oplus\, \mathfrak{h}$. These steps facilitate writing $\mathfrak{so}(2,3)$ Lie algebra in four avatars.
\end{enumerate}
The four inequivalent copies of $\mathfrak{sl}(2,{\mathbb R})$ embedded into $\mathfrak{so}(2,3)$ have $\mathfrak{h}$ as $\emptyset$, $\mathfrak{so}(1,1)$, $\mathfrak{so}(2)$ and $\mathfrak{sl}(2, {\mathbb R})$. Each of these leads to a chiral ${\cal W}$-algebra extension of $\mathfrak{so}(2,3)$. 
The resultant algebra corresponding to the case with $\mathfrak{h} = \emptyset$ can be identified to be the ${\cal W}(2,4)$ algebra found long ago \cite{Bouwknegt:1988sv}. Another, corresponding to $\mathfrak{h} = \mathfrak{sl}(2, {\mathbb R})$ is the chiral $\Lambda$-$\mathfrak{bms}_4$  of \cite{Romans:1990ta,Gupta:2022mdt}. We will complete this list by constructing the remaining two, with $\mathfrak{h} = \mathfrak{so}(1,1)$ and $\mathfrak{h} = \mathfrak{so}(2)$. The one with $\mathfrak{h} = \mathfrak{so}(1,1)$ may be called the conformal $\mathfrak{bms}_3$ whose semi-classical limit (that is, valid only for large values of central charges $c$ and $\kappa$) has been obtained from the asymptotic symmetries of 3$d$ conformal gravity \cite{Fuentealba:2020zkf}. The other $\mathcal W$-algebra with $\mathfrak h= \mathfrak{so}(2)$ is the close cousin of the one with $\mathfrak h=\mathfrak{so}(1,1)$ and has the same structure except for the changes accompanied by the Euclidean signature of the Killing form of current algebra $\mathfrak h$.

The rest of this paper is organised as follows. In section \ref{so23avatars} we provide a simple classification of all the inequivalent embedding of $\mathfrak{sl}(2,{\mathbb R})$ inside $\mathfrak{so}(2,3)$. This results in four different ways of writing the algebra $\mathfrak{so}(2,3)$ which makes the embedded $\mathfrak{sl}(2,{\mathbb R})$ manifest. In section \ref{sectionthree} we provide a derivation of two novel ${\cal W}$-algebra extensions of $\mathfrak{so}(2,3)$ completing the list of such algebras along with the two already known from \cite{Bouwknegt:1988sv} and \cite{Gupta:2022mdt}. In section \ref{dicussion} we conclude with a discussion and future directions. In Appendix \ref{imposingjacobi} we provide the details of the imposition of Jacobi Identities for one of our novel $\mathcal W$-algebra. In Appendix \ref{appendixA} we provide a similar derivation of chiral algebra extensions of $\mathfrak{so}(2,2)$ and show that they are isomorphic to the Brown-Henneaux algebra \cite{Brown:1986nw} and the CIG algebra of Polyakov \cite{Polyakov:1987zb} as realised in \cite{Avery:2013dja, Apolo:2014tua}. 

\section{Embedding $\mathfrak{sl}(2, {\mathbb R})$ in $\mathfrak{so}(2,3)$}
\label{so23avatars}
We are interested in promoting the algebra $\mathfrak{so}(2,3)$ to an infinite dimensional one that contains at least one Virasoro algebra. Each such Virasoro can in turn be thought of as an infinite dimensional extension of an $\mathfrak{sl}(2, {\mathbb R})$ subalgebra of $\mathfrak{so}(2,3)$. Therefore we would end up with a potentially distinct chiral ${\cal W}$-algebra extension of  $\mathfrak{so}(2,3)$ depending on which $\mathfrak{sl}(2, {\mathbb R}) \subset \mathfrak{so}(2,3)$ is promoted to a Virasoro. Therefore the first question to answer is: what are all the inequivalent embeddings of a copy of $\mathfrak{sl}(2, {\mathbb R})$ in the $\mathfrak{so}(2,3)$ algebra? 

This can be answered in any (faithful) matrix representation of the algebra $\mathfrak{so}(2,3)$. We choose to work with the matrix representation of $\mathfrak{so}(2,3)$ when the representation space is ${\mathbb R}^{2,3}$ (with the pseudo-Cartesian coordinates $( x_0, x_1, x_2, x_3, x_4)$ where $x^0, x^4$ are time-like and the rest space-like) on which the  $\mathfrak{so}(2,3)$ algebra elements generate linear homogeneous transformations $x^\mu \rightarrow {\Lambda^\mu}_\nu x^\nu$ that preserve the line element $ds^2 =  \eta_{\mu\nu} dx^\mu dx^\nu$, i.e., $\{ \Lambda \in M_5({\mathbb R}) : {\Lambda^\mu}_\alpha {\Lambda^\nu}_\beta \eta_{\mu\nu} = \eta_{\alpha\beta} ~ \& ~ {\rm det} \, \Lambda =1 \}$ where $\eta_{\mu\nu} = {\rm diag} (-1, 1, 1, 1, -1)$.\footnote{Another simple choice would have been to work with the $4 \times 4$ real matrix generators of the $\mathfrak{sp}(4, {\mathbb R})$ which is isomorphic to $\mathfrak{so}(2,3)$.} Here the algebra of $\mathfrak{so}(2,3)$ is realised in terms of $5 \times 5$ real matrices $M$ satisfying $M^{\rm T}\eta  + \eta M =0$. We will use the basis $L_{\mu\nu}$
\bea
\label{vectorofso23}
{(L_{\mu\nu})^\alpha}_\beta = \delta^\alpha_\mu \eta_{\nu\beta} - \delta^\alpha_\nu \eta_{\mu\beta}, ~~ ~~ \mu,\nu, \alpha, \beta \cdots \in \{ 0,1,2,3,4 \}
\eea
satisfying 
\bea
[L_{\mu\nu}, L_{\mu' \nu'} ] = \eta_{\mu\nu'} L_{\nu\mu'} + \eta_{\nu\mu'} L_{\mu\nu'} - \eta_{\mu\mu'} L_{\nu\nu'} - \eta_{\nu\nu'} L_{\mu\mu'} \, .
\eea
A general element in the Lie algebra is $M = \frac{1}{2} \omega^{\mu\nu} L_{\mu\nu}$ for $\omega^{\mu\nu} = - \omega^{\nu\mu} \in {\mathbb R}$, and in the space of $M$ we would like to identify choices of triplets $(L_0, L_{\pm 1})$ that form $\mathfrak{sl}(2, {\mathbb R})$ algebra 
\bea
\label{sl2alg}
[L_m, L_n] = (m-n) \, L_{m+n} \, .
\eea
We want all such possible choices that are not equivalent under the action of the group $O(2,3)$ which acts on $M$ as $M \rightarrow \Lambda^{-1} M \Lambda$ where the $5\times 5$ real matrices $\Lambda$ satisfy $\Lambda^{\!\rm T} \eta \Lambda = \eta$.\footnote{We do not require the $\det \Lambda =1$ condition to keep $\mathfrak{so}(2,3)$ invariant under $M \rightarrow \Lambda^{-1} M \Lambda$.} These realisations of $L_n$ will therefore be a five dimensional representation of $\mathfrak{sl}(2,{\mathbb R})$.

Before proceeding further let us recall some known facts about finite dimensional representations of $\mathfrak{sl}(2,{\mathbb R})$. The irreducible realisations of the $\mathfrak{sl}(2,{\mathbb R})$ are given in terms of $(2j+1) \times (2j+1)$ dimensional real matrices, one for each $j \in \frac{1}{2}{\mathbb N}$ (along with $j=0$, the singlet). Their matrix elements may be written as:
\bea
(L_0)_{qp} &=& p \, \delta_{qp}, ~~ (L_{-1})_{qp} = h(j,p) (p-j) \delta_{q,p+1}, \cr
(L_1)_{qp} &=& h(j,q)^{-1} (p+j) \delta_{q+1,p} \, , ~~~ {\rm for} ~~~ p,q \in \{-j, -j+1, \cdots, j-1, j\} \cr &&
\eea
for any arbitrary non-zero choice of $h(j,p)$ -- which can be fixed by appropriate actions of the group $GL(2, {\mathbb R})$. The simplest choices of $h(j,p)$ include: (i) $h(j,p)=1$ (or any other non-zero real number), (ii) $h(j,p) = \frac{1}{p-j}$,  or $ h(j,p) = p+j+1$ etc.\footnote{The choice $h(j,p) = \sqrt{\frac{p+j+1}{p-j}}$ is the one related to the discrete series unitary representations of $\mathfrak{sl}(2,{\mathbb R})$ when $j$ is analytically continued using the replacement $-j \rightarrow h$ for positive $h$ \cite{Jackiw:1990ka}. However this choice of course does not keep $L_{\pm 1}$ real.} In all these cases the $L_0$ is diagonal with rank $2j$ ($2j+1$) for $2j$ is even (odd), and with real eigenvalues. Its eigenvalues range over $-j, \cdots, j$. The only inputs we will take from here is that $L_0$ is diagonalisable with real eigenvalues, whereas $L_{\pm 1}$ are expected to be nilpotent. 

Next we seek all possible candidates in the algebra $\mathfrak{so}(2,3)$ in its vector representation (\ref{vectorofso23}) that can play the role of $L_0$. This requires us to consider the class of all $M$ that admit only real eigenvalues. We can take any such matrix $M$, using the freedom of acting with the group $O(2,3)$, to be a linear combination of any two commuting boost generators $ (L_{0a}, L_{b4})$ for $a\ne b$. There are six such choices which are all equivalent to each other under $O(2,3)$. So we choose:
\bea
L_0 
= \lambda \, L_{01} + \lambda' \, L_{34}
\eea
for $\lambda, \lambda' \in {\mathbb R}$ with real eigenvalues $(0, \pm \lambda, \pm \lambda')$. One can independently make the exchanges  $\lambda \leftrightarrow - \lambda$, $\lambda' \leftrightarrow - \lambda'$ and $\lambda \leftrightarrow \lambda'$ with appropriate $O(2,3)$ transformations.\footnote{These matrices are
%
$\Lambda = {\rm diag} (-1,1,-1,1,1), ~~ {\rm diag}(1,1,-1,1,-1), ~~ \Lambda_{mn} = \delta_{m+n}$ respectively.}
%
So we further choose $\lambda\ge \lambda' \ge 0$. Now taking this as the general form of $L_0$ we can seek candidate $L_{\pm 1}$ such that they form the algebra (\ref{sl2alg}) of $\mathfrak{sl}(2, {\mathbb R})$. There are, depending on the values of $(\lambda, \lambda')$, further $O(2,3)$ transformations that leave $L_0$ invariant and we can use these residual symmetries to put the candidate $L_{\pm 1}$ into their simplest forms. This exercise can be carried out systematically and straightforwardly. We find that, at the first instance, for the existence of non-trivial $L_{\pm 1}$ with $[L_0, L_{\pm 1} ] = \mp L_{\pm 1}$ requires that $(\lambda, \lambda')$ have to satisfy one of the following conditions: (i) $\lambda =1$, (ii) $\lambda'=1$, (iii) $ \lambda' =1-\lambda$ and (iv) $\lambda' =\lambda-1$. We can then impose the last condition $[L_1, L_{-1}] = 2L_0$ in each of these four cases. We simply list the results of this straightforward analysis:
\begin{enumerate}
\item $\lambda =1$ leads to three sub-cases (i) $\lambda' =1$, (ii) $0 < \lambda' < 1$ and (iii) $\lambda' =0$.
\begin{enumerate}
\item In the cases $\lambda' =1$ and $0 < \lambda' < 1$ there is no solution for $L_{\pm 1}$.
\item When $\lambda=1$ and $\lambda'=0$, with $L_0 = L_{01}$ the most general $L_{\pm 1}$ are:
\bea
(c_{12}^2 + c_{13}^2 - c_{04}^2) \, L_1 
&=& c_{12} (L_{12} - L_{02}) + c_{13}(L_{13}-L_{03}) + c_{04} (L_{14}-L_{04}) \cr
L_{-1} 
&=& c_{12} (L_{02} + L_{12}) + c_{13} (L_{03}+ L_{13}) + c_{04} (L_{04} + L_{14})   \cr &&
\eea
%
for $c_{12}^2 + c_{13}^2 - c_{04}^2 \ne 0$. It can be shown that the residual symmetries ($O(2,3)$ transformations that leave $L_0$ invariant) leave $c_{12}^2 + c_{13}^2 - c_{04}^2$ invariant up to a positive scale. Thus depending on the sign of $c_{12}^2 + c_{13}^2 - c_{04}^2$ we can take $(c_{12}, c_{13}, c_{04})$ to be either $(1,0,0)$ or $(0,0,1)$ and these two will be the only inequivalent choices in this case. 
\end{enumerate}
\item $\lambda'=1$ and $\lambda >1$ leads to two sub cases: (i) $\lambda =2$ and (ii) $\lambda \ne 2$.
\begin{enumerate}
\item For $\lambda =2$ and $\lambda'=1$ with $L_0 = 2 L_{01} + L_{34}$, we find:
%
\bea
\!\!\!\!\!\!\!\!\!\! b_{04} b_{23} L_1 
&=& b_{23} (L_{04}-L_{03} + L_{13}- L_{14}) + 3 b_{04} (L_{23} + L_{24}) \cr
L_{-1} 
&=& b_{04} (L_{03}+L_{04})+(L_{13}+L_{14}) + b_{23} (L_{23}-L_{24}) 
\eea
%
The residual symmetries can be used again to set $b_{23} = b_{04} =1$.
\item For $\lambda >1$ and $\lambda \ne 2$ leads to no solution.
\end{enumerate}
\item $\lambda' = 1- \lambda$ with $\frac{1}{2} \le \lambda < 1$ (since $\lambda' = 0$ is already considered) also leads to two sub-cases: (i) $\lambda = 1/2$ and (ii) $1/2 < \lambda <1$.
\begin{enumerate}
\item $\lambda = 1/2$ with $L_0 = \frac{1}{2} (L_{01} + L_{34})$ leads to a solution
\bea
4 \, a_{04} \, L_{-1} 
&=& - L_{03}+L_{04}- L_{13}+L_{14}\cr
~~ a_{04}^{-1} \,  L_1 
&=& L_{03}+L_{04} - L_{13}-L_{14}
\eea
again we can use the residual symmetries of $L_0$ to set $a_{04} = 1/2$.
\item $1/2 < \lambda <1$ -- leads to no solution.
\end{enumerate}
\item The final case is $\lambda' = \lambda -1$. This means $\lambda \ge 1$. We have already covered the cases of $\lambda =1$ and $\lambda = 2$.  So we restrict to $\lambda >1$ and not equal to $2$. And we can show that there is no solution in this case.
\end{enumerate}
Thus we have arrived at the result that there are precisely four inequivalent embeddings of $\mathfrak{sl}(2,{\mathbb R})$ into $\mathfrak{so}(2,3)$: (i) $(\lambda, \lambda') =(1,0)$ with $(c_{12}, c_{13}, c_{04}) = (1,0,0)$, (ii) $(\lambda, \lambda') =(1,0)$ with $(c_{12}, c_{13}, c_{04}) = (0,0,1)$, (iii) $(\lambda, \lambda') =(2,1)$, and (iv) $(\lambda, \lambda') =(1/2,1/2)$. As the next step, we will write down the algebra of $\mathfrak{so}(2,3)$ from the perspective of each of these embeddings. We choose to characterise these four avatars of $\mathfrak{so}(2,3)$ by their maximal subalgebras $\mathfrak{h}$ that commute with the embedded $\mathfrak{sl}(2, {\mathbb R})$.
\subsection{Four avatars of $\mathfrak{so}(2,3)$}
%
\subsubsection*{$\mathfrak{h} = \mathfrak{so}(1,1)$}
This case corresponds to $(\lambda, \lambda') =(1,0)$ \& $(c_{12}, c_{13}, c_{04}) = (1,0,0)$ with the generators
\bea
L_{-1} &=& L_{02}+L_{12}, ~~ L_0 = L_{01}, ~~ L_1 = - L_{02}+L_{12}, ~~ H = L_{34}, \cr
P^0_{-1} &=& -L_{04}-L_{14}, ~~ P^0_0 = L_{24}, ~~ P^0_1 = - L_{04}+L_{14} \, , \cr
P^1_{-1}&=& -L_{03}-L_{13}, ~~ P^0_0 = L_{23}, ~~ P^0_1 = - L_{03}+L_{13} \, ,\, 
\eea
satisfying
\bea
\label{avatar-one}
[L_m, L_n] &=& (m-n) L_{m+n} \, ,\cr
[L_m, P^a_n] &=& (m-n) P^a_{m+n} \, , \cr
[P^a_m, P^b_n] &=& -\eta^{ab} (m-n) L_{m+n} - \epsilon^{ab} \eta_{mn} \, H \, ,\cr
[H, P^a_m] &=& P^b_m \, {\epsilon_b}^a\, ,
\eea
for $m,n, \cdots \in \{0, \pm 1\}$ and $a,b, \cdots \in \{0,1\}$. Further $\epsilon^{01} = - \epsilon^{10} = 1$, $\eta_{ab} = {\rm diag} (-1,1) = \eta^{ab}$ and ${\epsilon_b}^a = \eta_{bc} \epsilon^{ca}$, and $\eta_{mn} = (3m^2-1)\delta_{m+n} = m^2+n^2-mn-1$. This avatar is related to the way $\mathfrak{so}(2,3)$ is realised as the conformal symmetry of ${\mathbb R}^{1,2}$. Clearly, in this case the subalgebra $\mathfrak{h}$ that commutes with the subalgebra $\mathfrak{sl}(2,{\mathbb R})$ is generated by the boost generator $H = L_{34}$ and hence is $\mathfrak{so}(1,1)$.
\subsubsection*{$\mathfrak{h} = \mathfrak{so}(2)$}
This case corresponds to $(\lambda, \lambda') =(1,0)$ \& $(c_{12}, c_{13}, c_{04}) = (0,0,1)$ with the generators
\bea
L_0 &=& L_{01}, ~~ L_1 = L_{04}-L_{14}, ~~ L_{-1} = L_{04}+L_{14}, ~~ R=L_{23}, ~~ \cr
P^1_{-1} &=& -L_{03}-L_{13}, ~~ P^1_0 = L_{34}, ~~ P^1_1 = L_{03}-L_{13}, \cr
P^2_{-1} &=& -L_{02}-L_{12}, ~~ P^2_0 = L_{24}, ~~ P^2_1 = L_{02}-L_{12} \, 
\eea
satisfying
\bea
\label{avatar-two}
[L_m, L_n] &=& (m-n) L_{m+n}, \cr
[L_m, P^a_n] &=& (m-n) P^a_{m+n}, \cr
[P^a_m, P^b_n] &=& \delta^{ab} (m-n) L_{m+n} + \epsilon^{ab} \eta_{mn} \, R \, ,\cr
[R, P^a_m] &=&  \epsilon^{ab} P^b_m \, ,
\eea
for $m,n, \cdots \in \{0, \pm 1\}$ and $a,b, \cdots \in \{1,2\}$. Further $\epsilon^{12} =1 = - \epsilon^{21}$. Contrary to the previous case ($\mathfrak{h} = \mathfrak{so}(1,1)$) the subalgebra $\mathfrak{h}$ is generated by the rotation generator $L_{23}$ and hence is $\mathfrak{so}(2)$. Thus, even though this case is very similar, it is not equivalent to the one with $\mathfrak{h} = \mathfrak{so}(1,1)$.
\subsubsection*{$\mathfrak{h} = \emptyset$}
This case corresponds to $(\lambda, \lambda') =(2,1)$. The generators are
\bea
L_0 &=& 2L_{01}+L_{34}, \cr
L_1 &=& - L_{03}+L_{04}+L_{13}-L_{14}+3L_{23}+3L_{24} \, , \cr
L_{-1} &=& L_{03}+L_{04}+L_{13}+L_{14}+L_{23}-L_{24} , \cr
W_{-3} &=& \beta (L_{03}-L_{04}+L_{13}-L_{14}), \cr
W_{-2} &=& \beta (L_{02} + L_{12}) , \cr
W_{-1} &=&  \beta \frac{3}{5} (-L_{03}-L_{04}-L_{13}- L_{14} + \frac{2}{3} (L_{23}-L_{24})) , \cr
W_0 &=& \beta \frac{3}{5} (-L_{01} + 2 L_{34}), \cr
W_1 &=& \beta \frac{3}{5} (L_{03}-L_{04}-L_{13}+ L_{14} + 2 (L_{23}+L_{24})) , \cr
W_2 &=& 3 \beta  (L_{02}-L_{12}), \cr
W_3 &=& 9 \beta  (-L_{03}- L_{04} + L_{13} + L_{14}) ,
\eea
satisfying
\bea
\label{avatar-three}
[L_m, L_n] &=& (m-n) \, L_{m+n} , \cr
[L_m, W_n] &=& (3m-n) \, W_{m+n} ,\cr
[W_m, W_n] &=& \frac{\beta^2}{100} \,  A(m,n) L_{m+n} + \frac{\beta}{10} \, B(m,n) W_{m+n},
\eea
where 
{\small 
\bea
\label{defAB}
A(m,n) &=&  (m-n) \Big(3(m^4+n^4)-2mn(m-n)^2 - 39 (m^2+n^2)+20 mn + 108 \Big), \cr
B(m,n) &=&  (m-n) (n^2+m^2-mn-7) \, .
\eea
}
We can fix $\beta$ to be any non-zero real number by a residual $O(2,3)$ transformation -- and we choose $\beta =10/\sqrt{1680}$ for later convenience. This avatar of $\mathfrak{so}(2,3)$ is special as the corresponding embedding of $\mathfrak{sl}(2, {\mathbb R})$ in it is what can be called the principal embedding -- as the rest of the generators form a single 5-dimensional irreducible representation of $\mathfrak{sl}(2, {\mathbb R})$.
\subsubsection*{$\mathfrak{h} = \mathfrak{sl}(2, {\mathbb R})$}
This case corresponds to $(\lambda, \lambda') = (1/2,1/2)$. The generators are
\bea
L_0 &=& \frac{1}{2} (L_{01} + L_{34}), ~~ L_1 = \frac{1}{2} (L_{03} + L_{04} - L_{13} -L_{14}), \cr
L_{-1} &=& \frac{1}{2} (-L_{03}+L_{04}- L_{13}+L_{14}), ~~ J_0 = \frac{1}{2} (L_{01} - L_{34}), \cr
J_1 &=& \frac{1}{2} (-L_{03}-L_{14}+L_{04}+L_{13}), ~~ J_{-1} = \frac{1}{2} (L_{03}+L_{14}+L_{04}+L_{13}) , \cr
G_{\frac{1}{2}, \frac{1}{2}} &=& \alpha (-L_{02}+L_{12}), ~~ G_{-\frac{1}{2}, \frac{1}{2}} = \alpha (-L_{23}-L_{24}), \cr G_{\frac{1}{2}, - \frac{1}{2}} &=& \alpha (L_{23}-L_{24}), ~~ G_{-\frac{1}{2}, - \frac{1}{2}} = \alpha (L_{02}+ L_{12}) ,
\eea
satisfying
\bea
\label{avatar-four}
[L_m, L_n] &=& (m-n) L_{m+n}, ~~ [J_a, J_b] = (a-b) J_{a+b}, \cr
[J_a, G_{sr}] &=& \frac{1}{2} (a-2s) \, G_{a+s, r}, ~~ [L_m, G_{sr}] = \frac{1}{2} (m-2r) \, G_{s, m+r}, \cr
[G_{sr}, G_{s'r'}] &=& - 2 \alpha^2 (\epsilon_{rr'} J_{s+s'} + \epsilon_{ss'} L_{r+r'}) \, 
\eea
where $m,n, \cdots \in \{0, \pm 1\}$, $a,b, \cdots \in \{0, \pm 1\}$ and $r, s, \cdots \in \{ \pm 1/2\}$, with $\epsilon_{-\frac{1}{2}, \frac{1}{2}} = 1$. Again we can choose $\alpha$ to be any non-zero real number and we fix it to be unity.
%
This completes the exercise of finding the inequivalent embeddings of $\mathfrak{sl}(2, {\mathbb R})$ inside $\mathfrak{so}(2,3)$ and writing the $\mathfrak{so}(2,3)$ algebra that makes the corresponding $\mathfrak{sl}(2, {\mathbb R})$ manifest. Each such embedding is based on the identification of a subalgebra $\mathfrak{sl}(2, {\mathbb R}) \oplus \mathfrak{h} \subset \mathfrak{so}(2,3)$ -- which are also in one-to-one correspondence with the maximal subalgebras of $\mathfrak{so}(2,3)$  listed in \cite{deGraaf:2019dni} that contain at least one copy of $\mathfrak{sl}(2, {\mathbb R})$ in them. 

Next we would like to promote each of these avatars (\ref{avatar-one}, \ref{avatar-two}, \ref{avatar-three}, \ref{avatar-four}) of $\mathfrak{so}(2,3)$ into an infinite dimensional algebra that consists of one copy of  Virasoro algebra. We will show in the next section that the first two avatars above admit minimal chiral ${\cal W}$-algebra extensions that are novel. The third avatar leads to the known ${\cal W}(2,4)$ algebra, and the extension of the fourth one is chiral $\Lambda$-$\mathfrak{bms}_4$ algebra. 
\section{The four ${\cal W}$-algebra extensions}
\label{sectionthree}
We now would like to find the minimal infinite dimensional extensions of the four avatars of $\mathfrak{so}(2,3)$ found in section \ref{so23avatars} such that they contain modes of a chiral stress tensor. We implement the following set of rules for this:
\begin{itemize}
\item For every copy of $\mathfrak{sl}(2, {\mathbb R})$ with generators $L_n$ (for $n = 0, \pm 1$)  we postulate a chiral quasi-primary stress tensor $T(z)$ with dimension $h=2$ and central charge $c$, and modes $L_n$ (for $n\in {\mathbb Z}$) with OPE,
\begin{align}
	T(z) T(w) \sim \frac{c/2}{(z-w)^4}+ \frac{2T(w)}{(z-w)^2} + \frac{\partial T(w)}{z-w} \, . \label{TT}
\end{align}
\item For the generators of the subalgebra $\mathfrak{h}$ we postulate a level-$\kappa$ current with $h=1$.
\item For every set of generators that form a non-trivial finite dimensional representation of $\mathfrak{sl}(2, {\mathbb R})$ of dimension $k$ we postulate a primary of dimension $h= \frac{1}{2} (k+1)$.
\item We write down the most general OPEs among the chiral operators as obtained above imposing the global symmetries at hand. 
\item Finally we fix the undermined coefficients in these OPEs using the OPE associativity - implemented in terms of commutators of the modes and their Jacobi identities explicitly. 
\end{itemize}
Following these rules each of the four avatars of $\mathfrak{so}(2,3)$ lead to the following sets of chiral operators. 
\begin{enumerate}
\item In the $\mathfrak{h} = \mathfrak{so}(1,1)$ case we have, along with $T(z)$, 
\begin{enumerate}
%
%
\item An $\mathfrak{so}(1,1)$ current $H(z)$  which is a conformal primary with $h=1$ and level-$\kappa$. 
\bea
T(z) H(w) &\sim & \frac{H(w)}{(z-w)^2} + \frac{ \partial H(w)}{z-w} ,\cr && \cr
H(z) H(w) &\sim & \frac{\kappa}{(z-w)^2}  \, . \label{TH}
\eea
\item A pair of chiral operators $P^a(z)$ that are conformal primaries with $h=2$ as well as a doublet of $\mathfrak{so}(1,1)$ current $H(z)$.\footnote{One could have postulated that $P^a(z)$ are only quasi-primaries at this stage. It turns out, however, that consistency of the final algebra with associativity does not allow central terms in $T(z) P^a(w)$ OPE and hence we preclude them.}
\bea
T(z) P^a(w) &\sim& \frac{2 P^a(w)}{(z-w)^2} + \frac{\partial P^a(w)}{z-w} \, ,
\cr && \cr
H(z) P^a(w) &\sim & \frac{ P^b(w) \, {\epsilon_b}^a\,}{z-w} \, . \label{TP}
\eea
This leaves the OPE $P^a(z) P^b(w)$ to be determined. We will refer to the resultant algebra to be ${\cal W}(2; 2^2, 1)$.
\end{enumerate}
\item In the $\mathfrak{h} = \mathfrak{so}(2)$ avatar we have, along with $T(z)$, 
\begin{enumerate}
\item An $\mathfrak{so}(2)$ current $R(z)$  which is a conformal primary with $h=1$ and level-$\kappa$.
\bea
T(z) R(w) &\sim & \frac{R(w)}{(z-w)^2} + \frac{ \partial R(w)}{z-w},  \cr && \cr
R(z) R(w) &\sim & \frac{\kappa}{(z-w)^2} \, .
\eea
\item A pair of chiral operators $P^a(z)$ that are conformal primaries with $h=2$ as well as a doublet of $\mathfrak{so}(2)$ current $R(z)$.
\bea
T(z) P^a(w) &\sim& \frac{2 P^a(w)}{(z-w)^2} + \frac{\partial P^a(w)}{z-w} \, ,
\cr && \cr
R(z) P^a(w) &\sim & \frac{ \epsilon^{ab} P^b(w) \,}{z-w}  \, .
\eea
The OPE $P^a(z) P^b(w)$ will be determined later. We will refer to the resultant algebra to be $\widetilde {\cal W}(2; 2^2, 1)$.
\end{enumerate}
\item In the case with $\mathfrak{h} = \emptyset$ we have a stress tensor $T(z)$ and
\begin{enumerate}
\item A chiral operator $W(z)$ that is a conformal primary of dimension $h=4$.
\bea
T(z) W(w) \sim \frac{4 W(w)}{(z-w)^2} + \frac{ \partial W(w)}{z-w} \label{TW OPE}
\eea
The $W(z) W(w)$ will be determined later. The resultant algebra is shown to be ${\cal W}(2,4)$ algebra of \cite{Bouwknegt:1988sv}.
\end{enumerate}
\item In the case with $\mathfrak{h} = \mathfrak{sl}(2, {\mathbb R})$ we have a $T(z)$ and
\begin{enumerate}
\item A triplet $J_a(z)$ with $a=0, \pm 1$, which are conformal primaries with $h=1$ 
\bea
 T(z) J_a(w) &\sim& (z-w)^{-2} J_a(w) + (z-w)^{-1} \partial J_a(w) \, ,
\eea
and among themselves form a level-$\kappa$ $\mathfrak{sl}(2, {\mathbb R})$ algebra,
\bea
J_a(z) J_b(w) &\sim& -\frac{\kappa}{2} \eta_{ab} (z-w)^{-2} + (z-w)^{-1}{f_{ab}}^c J_c(w) \, .
\eea
\item A pair $G_s(z)$ for $s=\pm 1/2$ that are conformal primaries with $h=3/2$ 
\bea
T(z) G_s(w) &\sim& (3/2) (z-w)^{-2} G_s(w) + (z-w)^{-1} \partial G_s(w)
\eea
as well as a doublet of $\mathfrak{sl}(2, {\mathbb R})$ current algebra
\bea
J_a(z) G_s(w) &\sim& (z-w)^{-1} G_{s'}(w) {(\lambda_a)^{s'}}_s 
\eea
\end{enumerate}
where ${(\lambda_a)^s}_{s'} = \frac{1}{2} (a-2s') \delta^{s}_{a+s'} $. This case was already dealt with by us in \cite{Gupta:2022mdt} and the OPE $G_s(z) G_{s'}(w)$ was determined. The resultant algebra is called chiral $\Lambda$-$\mathfrak{bms}_4$ algebra .
\end{enumerate}
We will now turn to obtaining the underdetermined OPEs in each of these cases. These will be determined using the standard tools of 2d CFT where we impose Jacobi Identities on the commutators of the modes of chiral operators. For the convenience of the readers, we will provide the essential details of the derivation of first of these algebras and the results for the remaining three.
\subsection{${\cal W}(2; 2^2,1)$}
In this case the only undetermined OPE is that of $P^a(z) P^b(w)$ and following the conformal invariance one expects that all quasi-primaries of dimension up to three and their global descendants can appear in the singular terms on the rhs. These quasi-primaries can also be composites (normal ordered products) of simple (quasi-) primaries $(T(z), P^a(z), H(z))$. One can construct the necessary quasi-primaries using the OPEs \eqref{TT},\eqref{TH}, \eqref{TP}.
The relevant composite quasi-primaries are as follows.
\begin{itemize}
\item At $h=2$ we have only one normal-ordered composite quasi-primary $(H^2)(z)$ with
\bea
T(z) (H^2)(w) \sim \frac{\kappa}{(z-w)^4} + \frac{2}{(z-w)^2}(H^2)(w) + \frac{1}{(z-w)}\partial (H^2)(w) \, .
\eea
The corresponding Sugawara stress tensor $\frac{1}{2\kappa} (H^2)(z)$ has $c = 1$.
\item At $h=3$ we have two composite scalar quasi-primaries
\bea
\label{defLS}
\Lambda(z) &:=& (TH)(z) - \frac{1}{2} \partial^2 H(z), \cr
\Sigma(z) &:=& \frac{1}{2\kappa}((H^2)H)(z) - \frac{1}{2} \partial^2 H(z).
\eea
Their OPEs with $T(z)$ are given by
\bea
T(z) \Lambda(w) &\sim& \frac{2+c}{2(z-w)^4} H(w) + \frac{3}{(z-w)^2} \Lambda(w) + \frac{1}{(z-w)}\partial \Lambda(w) \, ,\cr && \cr
T(z) \Sigma(w) &\sim& \frac{3}{2(z-w)^4} H(w) + \frac{3}{(z-w)^2} \Sigma(w) + \frac{1}{(z-w)}\partial \Sigma(w) \, .
\eea
Note that the linear combination $3 \, \Lambda (z) - (2+c) \, \Sigma(z)$ is a primary.
\item At $h=3$ we have a doublet of primaries (not just quasi-primaries)
\bea
\label{HPa}
[P^aH](z) := (P^aH)(z) + \frac{3}{4} {\epsilon_b}^a \partial P^b =(HP^a)(z)- \frac{1}{4} {\epsilon_b}^a \partial P^b
\eea
\end{itemize}
This is the full list of composite quasi-primaries that are covariant under $\mathfrak{so}(1,1)$ and dimension $h\le 3$. In particular, let us note that there is no rank-2 symmetric traceless tensor of $\mathfrak{so}(1,1)$ with $h \le 3$. There are other quasi-primaries of higher dimensions that we will introduce as and when required. 

We now make use of the global symmetries to constrain the OPE $P^a(z) P^b(w)$. 
\begin{itemize}
\item Since the two operators involved carry vector indices $a,b, \cdots$ of $\mathfrak{so}(1,1)$ their OPE should respect this symmetry. This means that the rhs can only involve representations of $\mathfrak{so}(1,1)$ that can appear in the tensor product of two vectors. 
\item There are in general four independent components in a rank-2 tensor of $\mathfrak{so}(1,1)$. Arranging them into covariant irreducible objects, there are 
\begin{enumerate}
\item two singlets - with the CG coefficients being the symmetric tensor $\eta^{ab}$ and the anti-symmetric tensor $\epsilon^{ab}$. These can be multiplied by $\mathfrak{so}(1,1)$ invariants. 
\item a symmetric traceless rank-2 tensor with two independent components. 
\end{enumerate}
%
\item One consequence of these tensor structures is that we cannot have a covariant rank-1 object (vector) on the rhs and that rules out (\ref{HPa}).  Also the fact that there are no composite operators with $h\le 3$ that are rank-2 traceless symmetric tensors leaves us with, in the increasing order of conformal dimensions $\{I, H(z), T(z), (H^2)(z), \Lambda, \Sigma\}$ that can appear on the rhs. 
\item Another fact is that the OPE is expected to be invariant under the simultaneous exchange  $(a, z) \leftrightarrow (b,w)$.  It is easy to see that this requires that the terms containing $\eta^{ab}$ should have even dimensional operators, namely, $(I, T(z))$ (and their descendants) and terms containing $\epsilon^{ab}$ should have odd dimensional operators, namely, ($H(z), \Lambda(z), \Sigma(z)$) and their descendants. 
\item The relative coefficients of a quasi-primary and its global conformal descendants are fixed by the global conformal invariance. This leaves only six real numbers to be undetermined. 
\end{itemize}
Thus we arrive at the following ansatz:
%
\bea
P^a(z) P^b(w) &\sim& \eta^{ab} \Big[\alpha \, (z-w)^{-4} \,  I + \beta\, (z-w)^{-2} \,  \Big( 2 T (w)+(z-w) \, \partial T(w) \Big) \cr
&& \cr
&& ~~~~~~~~~~ + \gamma (z-w)^{-2} \, \Big( 2(H^2)(w)+ (z-w) \partial (H^2)(w) \Big)  \Big] \cr
&& \cr
&& \!\!\!\!\! + \, \epsilon^{ab} \Big[ \delta \, (z-w)^{-3}  \Big( 6 H(w) +3 \, (z-w) \partial H(w) +  \, (z-w)^2 \, \partial^2 H(w) \Big) \cr && \cr
&& ~~~~~~~~~~ + \sigma \, (z-w)^{-1} \Lambda(w) + \omega \, (z-w)^{-1} \Sigma(w) \Big] \cr \label{PP}
&&
\eea
%
The undetermined real numbers $(\alpha, \beta, \gamma, \delta, \sigma, \omega)$ (that can depend on the central charges $(c, \kappa)$) are determined by imposing the OPE associativity using Jacobi identities in terms of modes. The details of this calculation are provided in the appendix \ref{imposingjacobi}.

The full set of conditions on the undetermined real numbers obtained from the Jacobi identities are
\begin{align}
\label{conds11}
&2\alpha-c\beta-2\kappa \, \gamma = 0, ~~~ 24 \delta - (2+c)\sigma - 3\omega = 0\,,~~ 
\beta +3 \delta +2\kappa  \gamma = 0 \cr 
&\alpha + 6 \kappa \, \delta = 0 \, ~~2\beta +\kappa \, \sigma = 0, ~~~
2\gamma + \sigma + \frac{3}{2} \omega = 0 \, ~~	4\kappa(\gamma-\sigma)-3\omega = 0 \cr
&4\kappa(2\beta+\gamma-2\delta+\sigma)+\omega =0 
\end{align}
Out of these eight only six equations are linearly independent. These
equations admit a non-trivial two-parameter solution.\footnote{There is of course the trivial solution $\alpha=\beta=\gamma=\delta = \sigma = \omega =0$, for arbitrary $c$ and  $\kappa$.}
 Taking these parameters to be $(\sigma, \kappa)$ we find:
\bea
\label{cbms3-result}
\alpha &=& \frac{3 \kappa^2(\kappa-1)}{\kappa+1} \sigma, ~~ \beta = - \frac{\kappa}{2} \sigma, \cr
\gamma &=& \frac{2\kappa-1}{2(\kappa+1)} \sigma, ~~ \delta  = - \frac{\kappa( \kappa-1)}{2 (\kappa+1)} \sigma,\cr
\omega &=& - \frac{2\kappa}{\kappa+1}\sigma, ~~ c = - \frac{2 \, (6\kappa^2-8\kappa+1)}{\kappa+1} \, .
\eea
This completes our derivation of the algebra ${\cal W}(2; 2^2,1)$. In the large-$\kappa$ limit these become
\bea
\label{largek}
\alpha &\rightarrow & 3 \kappa^2  \sigma, ~~ \beta \rightarrow - \frac{\kappa}{2} \sigma, ~~ \gamma \rightarrow  \sigma, \cr
\delta  &\rightarrow& - \frac{\kappa}{2 } \sigma, ~~~
\omega \rightarrow - 2 \, \sigma, ~~ c \rightarrow -12 \, \kappa \, .
\eea
By further taking $\sigma \rightarrow \frac{2}{\kappa}$ it can be seen that this ${\cal W}$ algebra, when restricted to the modes of $\mathfrak{so}(2,3)$ is a deformation of (\ref{avatar-one}) around the $\kappa \rightarrow \infty$ limit.

\subsubsection*{Relation between ${\cal W}(2; 2^2,1)$ and conformal $\mathfrak{bms}_3$ of \cite{Fuentealba:2020zkf}}
Before we turn to the next case let us briefly demonstrate that the conformal $\mathfrak{bms}_3$ algebra of \cite{Fuentealba:2020zkf} arises in the semiclassical limit of our ${\cal W}(2; 2^2,1)$.
For this we define
	\begin{align}
			\mathcal{P}_{m}= P^{0}_{m}+P^1_{m}\,,~\mathcal{K}_{m}=P^0_{m}-P^1_m
	\end{align}
and re-write the commutator (\ref{ppcomm}) in terms of $\mathcal{P}_m$ and $\mathcal{K}_n$. This results in, for instance, the commutator $\left[\mathcal{P}_m\,,\mathcal{K}_n\right]$
\begin{align}
\label{cbms3}
&\left[\mathcal{P}_m\,,\mathcal{K}_n\right]=-\frac{\alpha}{3}\,m(m^2-1)\delta_{m+n,0}-2(m-n)\left(\beta\,L_{m+n}+\gamma \, (H^2)_{m+n} \right) \nonumber \\& - 2\,\delta \, (m^2+n^2-mn -1) \, H_{m+n} -2\, \sigma \, \Lambda_{m+n} -2\, \omega \, \Sigma_{m+n} \Big]
\end{align}
%
Next we take the large-$\kappa$ limit which amounts to using (\ref{largek}). 
If one further replaces $c\rightarrow 12\, \tilde c$, $\kappa \rightarrow -\tilde c$ and $\sigma \rightarrow 2/{\tilde c}$ one obtains the conformal $\mathfrak{bms}_3$ algebra as written down in \cite{Fuentealba:2020zkf} after the following  identifications
\begin{align}
	 H_p \rightarrow i \mathcal{D}_p\,,~~ L_n \rightarrow \mathcal{J}_n.
\end{align} 
The normal ordered operators in this large-$\kappa$ limit after regularisation become
\begin{align}
&&	(H^2)_{m} \rightarrow - \sum_p \mathcal{D}_{m-p}\,\mathcal{D}_{p}\,,~~
	\Lambda_m \rightarrow i \sum_{p}\mathcal{J}_{m-p}\,\mathcal{D}_{p} \,,~~
\Sigma_m \rightarrow -\frac{i}{2 \tilde c}\sum_{n,l}\mathcal{D}_{m-n-l}\mathcal{D}_{n}\mathcal{D}_l
\end{align}
completing the identification with the non-linear terms of (\ref{cbms3}) with the expressions in (2.2--2.4) of \cite{Fuentealba:2020zkf}. This makes it clear that the conformal $\mathfrak{bms}_3$ algebra of \cite{Fuentealba:2020zkf} is only valid for large values of their central charge.

\subsection{$\widetilde{\cal W}(2; 2^2,1)$}
Now we turn to the second way the $\mathfrak{sl}(2,\mathbb{R})$ is embedded in $\mathfrak{so}(2,3)$ where $P^a$ instead of  being a vector in $\mathfrak{so}(1,1)$ algebra is vector under $\mathfrak{so}(2)$ algebra. This gives rise to a different $\mathcal{W}$-algebra extension which closely resembles ${\cal W}(2; 2^2,1)$ algebra found in previous section. The structure of all the quasi-primaries and primaries are expected to remain the same with some minor modifications (such as $H(z) \rightarrow R(z)$) and therefore we provide only the result here. 
\begin{enumerate}
	\item {\bf Virasoro Algebra}:
	\bea
	[L_m, L_n] = (m-n) \, L_{m+n} + \frac{c}{12} m (m^2-1) \, \delta_{m+n,0}
	\eea
	\item {\bf Virasoro Primaries}
	\bea
	[L_m, P^a_n] = (m-n) \, P^a_{m+n}, ~~ [L_m, R_n] = -n \, R_{m+n}
	\eea
	\item {\bf $\mathfrak{so}(2)_\kappa$ Current Algebra}
	\bea
	[R_m, R_n] = \kappa \, m \, \delta_{m+n, 0}
	\eea
	\item {\bf Current algebra primaries}
	\bea
	[R_m, P^a_n] = {\epsilon^{ab}} P^b_{m+n} 
	\eea
	\item {\bf Commutator $[P^a_m, P^b_n]$}
	%
		\bea
		[P^a_m, P^b_n] 
		&=& \delta^{ab} \left[ \frac{\alpha}{6} m (m^2-1) \delta_{m+n,0} +  (m-n) \, \Big( \beta \, L_{m+n} + \gamma \, (R^2)_{m+n} \Big) \right] \cr
		&& \cr
		&& + \epsilon^{ab} \Big[ \delta \, (m^2+n^2-mn -1) \, R_{m+n} + \sigma \, \Lambda_{m+n} + \omega \, \Sigma_{m+n} \Big] \cr &&
		\eea
		%
%
where $\Lambda_m $ and $\Sigma_m$ are the modes of those in (\ref{defLS}) with $H$ replaced by $R$. As before we impose Jacobi identities to solve for unknown parameters in the $\left[P^a_m,P^b_n\right]$ commutator.
From Jacobi Identity $(TPP)$ we get the following equations
\begin{align}
\label{avatar2-eqs1}
	2\alpha-c\,\beta -2 \gamma \,\kappa=0, ~~
	24\,\delta-(2+c)\,\sigma-3\,\omega=0
\end{align}
The $(RPP)$ Jacobi Identity demands
\begin{align}
\label{avatar2-eqs2}
	\beta+3\,\delta+2\,\gamma\,\kappa&=0, ~~
	\alpha-6\,\kappa\,\delta=0, \nonumber\\
	2\,\beta-\kappa\,\sigma&=0, ~~
	2\gamma-\sigma-\frac{3}{2}\,\omega=0	.
\end{align}
In imposing the Jacobi Identity $(PPP)$ we find similar structures that we saw before resulting in the following equations,
\begin{align}
\label{avatar2-eqs3}
	4\,\kappa \,\left(\gamma+\sigma\right)-3\,\omega&=0, ~~
	4\,\kappa\,\left(2\beta-\gamma-2\,\delta+\sigma\right)-\omega=0 .
\end{align}
These equations (\ref{avatar2-eqs1}, \ref{avatar2-eqs2}, \ref{avatar2-eqs3}) with replacements $\gamma \rightarrow -\gamma$ and $\kappa \rightarrow -\kappa$ match with the equations (\ref{conds11}, \ref{conds12}, \ref{conds13}, \ref{conds14}) obtained in $\mathfrak{so}(1,1)$ case. Hence we obtain the new algebra $\widetilde{\cal W}(2; 2^2, 1)$ where the parameters are given by (\ref{cbms3-result}) after the replacements $\gamma \rightarrow -\gamma$ and $\kappa \rightarrow -\kappa$.
\subsection{$\mathcal{W}$(2, 4)}
This chiral $\mathcal W$-algebra extension is based on writing $\mathfrak{so}(2,3)$-algebra as \eqref{avatar-three}. The OPE \eqref{TW OPE} leads to the following commutation relations,
\bea
 \left[L_m,W_n\right]=(3m-n)\,W_{m+n}
\eea
Using the method that was spelt out earlier we start with the following ansatz of $\left[W_m,W_n\right]$, 
\begin{align}
\label{w24ansatz}
	\left[W_m,W_n\right]&=\,\alpha\,\tilde{\alpha}(m,n)+\beta\, A(m,n)\,L_{m+n}+ B(m,n) \, \left( \gamma\, W_{m+n}+\delta\,\Lambda_{m+n} \right) \nonumber \\&+(m-n)\,\Big[\omega\,\Delta_{m+n}+ \tau\,\Gamma_{m+n}+ \sigma\,\Omega_{m+n}\Big]
\end{align}
where,
\begin{align}
	\tilde{\alpha}(m,n)&=\frac{m\,(m^6-14 m^4+49 m^2-36)}{5040}\,\delta_{m+n,0}
\end{align}
The modes  $\Lambda_{m}$ are associated with the $h=4$ quasi-primary 
\begin{align}
	\Lambda(z)&=\left(TT\right)(z)-\frac{3}{10}\partial^2 T(z). 
\end{align}
The modes $\Delta_{m},\Gamma_{m}\,,\Omega_{m}$ are those of the following three $h=6$ quasi-primaries:
\bea
	\Delta(z)&=&\left(T\,\Lambda\right)(z)-\frac{1}{6}\,\partial^2\Lambda(z), \cr
	\Gamma(z)&=&\left(\partial^2TT\right)(z)-\partial(\partial TT)(z)+\frac{2}{9}\,\partial^2(TT)(z)-\frac{1}{42}\partial^4 T(z), \cr
	\Omega(z)&=&(TW)(z)-\frac{1}{6}\partial^2W(z) \, .
\eea
In order to obtain equations constraining the parameters $\{\alpha,\beta,\gamma,\delta,\omega,\sigma,\tau\}$ in (\ref{w24ansatz}) we will impose Jacobi identities as for the earlier cases. From the Jacobi identity of $(TWW)$ one gets the following equations,
\begin{align}
	&\alpha-420\,c\,\beta=0,~~1960\,\beta-(22+5c)\,\delta=0,~~14\,\gamma-\frac{1}{12}(24+c)\,\sigma=0 , \nonumber\\&14\delta-\frac{1}{180}(160\,\tau+3\,\omega(15c+164))=0,  \nonumber\\ &737100\,\beta-315(22+5c)\,\delta-5(29+70c)\,\kappa-42(22+5c)\,\omega=0 \, .
\end{align}
From imposing $(WWW)$ Jacobi Identity one gets,
\begin{align}
	18900\,\beta-8820\,\delta-315\,\gamma\,(60\,\gamma-7\,\sigma)+95\,\tau+1029\,\omega=0
\end{align}
The solution to these equations are as follows,
\begin{align}
	&\alpha=420\,c\,\beta\,,~~\delta=\frac{1960\,\beta}{22+5c}\,,~~\omega=\frac{20160(13+72\,c)\,\beta}{70c^3+953 c^2+2498\,c-1496} \, ,\nonumber\\
	\nonumber\\
	&\tau=\frac{252\,(-524+19c)\,\beta}{14c^2+129c-68}\,,~~\sigma=\frac{168\,\gamma}{24+c}\,,~~\gamma^2=\frac{70\,\beta(c^3-148c^2-3932c+4704)}{70c^3+953c^2+2498c-1496}
\end{align} 
where $\beta$ is the free parameter, the value of which can be fixed.
Choosing $\beta=\frac{1}{1680}$ one can check that the above algebra is ${\cal W}(2,4)$ algebra as was written down in \cite{Bouwknegt:1988sv}.
\end{enumerate}
\subsection{Chiral $\Lambda$-$\mathfrak{bms}_4$ }
This case was dealt with in detail in \cite{Gupta:2022mdt} motivated from the algebra derived in the context of symmetries responsible for soft theorems in the context of MHV graviton amplitudes in 4$d$ flat spacetimes. Here we simply quote the result for the only non-trivial OPE
\bea
\label{ope-three}
G_s(z) G_{s'}(w) &=&  2 \alpha \, \epsilon_{ss'} (z-w)^{-3} + \delta \, (z-w)^{-2} \, {(\lambda^a)}_{ss'} \left[ 2 \, J_a(w) + (z-w) \, {J_a}'(w) \right] \cr && \cr
&& \!\!\!\!\! + \beta \, \epsilon_{ss'} (z-w)^{-1} \,   T(w) + \gamma \, (z-w)^{-1} \, \epsilon_{ss'} (J^2)(w) \, ,
\eea
which in terms of modes reads
\bea
&& \!\!\!\!\! [G_{s,r}, \! G_{s',r'}] \!= \!\epsilon_{ss'} \!\! \left[\alpha \! \left(r^2 - \frac{1}{4} \right) \! \delta_{r+r',0} \! + \! \beta \,  L_{r+r'}  \! + \! \gamma \, (J^2)_{r+r'} \right] \!\!+ \!\delta (r-r')  J_{a, r+r'}  {(\lambda^a)}_{ss'} \cr &&
\eea
where $(J^2)_n$ are the modes of the normal ordered quasi-primary $\eta^{ab} (J_aJ_b)(z)$:
\bea
(J^2)_n = \eta^{ab} \left[\sum_{k > -1} J_{b,n-k} J_{a,k} + \sum_{k \le -1}  J_{a,k} J_{b, n-k} \right] \nonumber
\eea
The result of imposition of the Jacobi identities is the following set of constraints on the parameters $(\alpha,\beta,\gamma,\delta, c, \kappa)$
\bea
\alpha -\frac{c}{6} \beta +\gamma \, \frac{\kappa}{2} =0, ~~ \alpha -\delta \, \frac{\kappa}{2} =0, ~~ \beta - \gamma \, (\kappa +2) - \frac{1}{2} \delta =0, ~~ \beta-\frac{\gamma-\delta}{2} =0 ~ .\nonumber \\
\eea
whose solution exists provided
\bea
\label{final-result-one}
c = - \frac{6 \kappa \, (1+ 2 \kappa)}{5+2\kappa} ~ .
\eea
When this holds, for generic values of $\kappa$ ($\ne -5/2$) the solution can be written (taking $\gamma$ to be the independent variable) as
\bea
\label{final-result-two}
\alpha = - \frac{1}{4} \gamma \kappa (3+2\kappa), ~~ \beta =  \frac{1}{4}\gamma(5+2\kappa), ~~ \delta = -\frac{1}{2} \gamma (3+2\kappa) \, . 
\eea
This is the final result valid for generic ($\kappa \ne -5/2$) values of $\kappa$ \cite{Gupta:2022mdt}. 
\vskip .2cm
This completes our exercise of finding all minimal chiral ${\cal W}$-algebra extensions of $\mathfrak{so}(2,3)$.

\section{Discussion}
\label{dicussion}
In this paper we have addressed the question:
\begin{center}
What are all the chiral ${\cal W}$-algebra extensions of $\mathfrak{so}(2,3)$? 
\end{center}
As the first step we have shown that there are exactly four ways to embed a copy of $\mathfrak{sl}(2, {\mathbb R})$ into $\mathfrak{so}(2,3)$ algebra. Each such embedding results in a distinct chiral ${\cal W}$-algebra extension and we have constructed two new ones, and recovered the other two that are known from \cite{Bouwknegt:1988sv, Romans:1990ta,Gupta:2022mdt} -- thus completing the list of such chiral extension of $\mathfrak{so}(2,3)$. One of our new algebras (namely ${\cal W}(2; 2^2,1)$ with $\mathfrak{h} = \mathfrak{so}(1,1)$) is the conformal $\mathfrak{bms}_3$ whose semi-classical version was first found in \cite{Fuentealba:2020zkf}.
\vskip .2cm
\noindent{We offer a few comments below:}
\begin{itemize}
\item Just as the ${\cal W}_3$ algebra of Zamalodchikov \cite{Zamolodchikov:1985wn} and the $W_3^{(2)}$ algebra of Bershadsky-Polyakov \cite{Polyakov:1989dm, Bershadsky:1990bg} could be realised \cite{Ammon:2011nk,Campoleoni:2010zq, Henneaux:2010xg, Campoleoni:2011hg} in the Chern-Simons theory with $\mathfrak{sl}(3, {\mathbb R})$ gauge algebra, each of the ${\cal W}$-algebras considered in this paper can also be realised as the asymptotic symmetry algebra of 3$d$ conformal gravity with appropriate boundary conditions. In \cite{Fuentealba:2020zkf}, the semi-classical limit of $\mathcal W(2;2^2,1)$ was already realised as the asymptotic symmetry algebra of $3d$ conformal gravity. We will present the realisation of all the others from $3d$ conformal gravity in an upcoming work.
\item Each extension we considered depends on the list of simple primaries \cite{Blumenhagen:1990jv} that are declared.  As we have seen there are primaries of higher dimensions in each chiral algebra and we could probably have worked with other sets of primaries as the simple ones and these might lead to other extensions $\mathfrak{so}(2,3)$. The choices we made of sets of simple primaries are with the minimal conformal dimensions, and therefore the extensions we considered are in a sense minimal. It will be interesting to construct the non-minimal algebras as well (if they exist) and study them. 
\item As we have seen, in each of the four chiral ${\cal W}$-algebras we constructed, the constraints from Jacobi identities admit trivial solutions as well (where all the parameters vanish and central charges are arbitrary, for instance, as mentioned in footnote (7) on page 17). These trivial solutions do not correspond to extensions of $\mathfrak{so}(2,3)$ but instead of different contractions of it. For instance in the case of chiral $\Lambda$-$\mathfrak{bms}_4$ of \cite{Gupta:2022mdt}, it corresponds to chiral $\mathfrak{bms}_4$ algebra, a chiral conformal algebra extension of $\mathfrak{iso}(1,3)$ ($\mathfrak{iso}(2,2)$) -- which is the isometry algebra of ${\mathbb R}^{1,3}$ (${\mathbb R}^{2,2}$). In the other three cases the analogous chiral conformal extensions are of other 10-dimensional algebras that are still contractions of $\mathfrak{so}(2,3)$. It will be interesting to identify spacetimes which admit these contraction of $\mathfrak{so}(2,3)$ as their symmetry algebras and ask if the corresponding infinite dimensional chiral conformal algebra extensions have any important role in them.

\item Even though we aimed to construct chiral ${\cal W}$-algebras it does not mean that we miss out on algebras that are not chiral. As an example we show in the appendix \ref{appendixA} that if we repeat the same exercise for $\mathfrak{so}(2,2)$ we end up with two chiral infinite dimensional algebras. One of them is indeed a chiral algebra which is an extension of CIG algebra of Polyakov \cite{Polyakov:1987zb} as derived by \cite{Avery:2013dja}. The other can be identified with the non-chiral Brown-Henneaux algebra though seen from the perspective of the diagonal Virasoro. It will be interesting to explore this question for our extensions of $\mathfrak{so}(2,3)$ as well.
%
\end{itemize}
The next step in analysing the algebras of this paper and their utility in the context of holography would be to construct their representations in terms of generalised primary operators and study the Ward identities of their correlation functions. Once that is achieved it will be important to provide prescription to compute them holographically, either in the 3$d$ conformal gravity theories or $AdS_4$-gravity theories. We hope to address these in the near future. 

\section*{Acknowledgements} We thank Suresh Govindarajan, Bala Sathiapalan and Adarsh Sudhakar for helpful conversations. Thanks are also due to the participants of the 5th edition of Chennai Strings Meeting (February 2023) for helpful comments on parts of this work presented there. 
\appendix
\section{Imposition of Jacobi identities for $\mathcal{W}(2;2^2,1)$}
\label{imposingjacobi}
In order to impose Jacobi identities we need the mode expansions of quasi-primaries as well as composite primaries. Using the definition from \cite{DiFrancesco:1997nk, Blumenhagen:2009zz}
\bea
(AB)_m = \sum_{n \le -h_A} A_n B_{m-n} + \sum_{n > -h_A} B_{m-n} A_n \, .
\eea
for the modes of normal ordered products we can write
\bea
(H^2)_n = \sum_{k >-1} H_{n-k} H_k + \sum_{k\le -1} H_k H_{n-k} \, . 
\eea
Similarly we have $\Lambda(z) = \sum_{n \in {\mathbb Z}} \frac{{\Lambda}_n}{z^{n+3}}$ where
\bea
\Lambda_n = \sum_{k >-2} H_{n-k} L_k + \sum_{k\le -2} L_k H_{n-k} - \frac{1}{2} (n+1)(n+2) H_n
\eea
and $\Sigma(z) = \sum_{n \in {\mathbb Z}} \frac{{\Sigma}_n}{z^{n+3}}$ where
\bea
	\Sigma_n = \sum_{k >-2} H_{n-k} \tilde L_k + \sum_{k\le -2} \tilde L_k H_{n-k} - \frac{1}{2} (n+1)(n+2) H_n
\eea
with $\tilde L_n = \frac{1}{2\kappa} (H^2)_n$. 
Let us convert all the OPEs \eqref{TT},\eqref{TP}, \eqref{TH} and \eqref{PP} into commutators of modes defined by
\bea
L_n = \oint_0 \frac{dz}{2\pi i} z^{n+1} T(z), ~~ P^a_n = \oint_0 \frac{dz}{2\pi i} z^{n+1} P^a(z), ~~ H_n = \oint_0 \frac{dz}{2\pi i} z^n \, H(z)
\eea
%
%
resulting in
\begin{enumerate}
	\item {\bf Virasoro Algebra}:
	\bea
		[L_m, L_n] = (m-n) \, L_{m+n} + \frac{c}{12} m (m^2-1) \, \delta_{m+n,0}
	\eea
	\item {\bf Virasoro Primaries}
	\bea
		[L_m, P^a_n] = (m-n) \, P^a_{m+n}, ~~ [L_m, H_n] = -n \, H_{m+n}
	\eea
	\item {\bf $\mathfrak{so}(1,1)_\kappa$ Current Algebra}
	\bea
		[H_m, H_n] = \kappa \, m \, \delta_{m+n, 0}
	\eea
	\item {\bf Current algebra primaries}
	\bea
		[H_m, P^a_n] =  P^b_{m+n} \, {\epsilon_b}^a
	\eea
	\item {\bf Commutator $[P^a_m, P^b_n]$}
	%
		\bea
		\label{ppcomm}
		[P^a_m, P^b_n] 
		&=& \eta^{ab} \left[ \frac{\alpha}{6} m (m^2-1) \delta_{m+n,0} +  (m-n) \, \Big( \beta \, L_{m+n} + \gamma \, (H^2)_{m+n} \Big) \right] \cr
		&& \cr
		&& + \epsilon^{ab} \Big[ \delta \, (m^2+n^2-mn -1) \, H_{m+n} + \sigma \, \Lambda_{m+n} + \omega \, \Sigma_{m+n} \Big] \cr &&
		\eea
		%
\end{enumerate}
Now we determine the unknowns $(\alpha,\beta, \gamma, \delta, \sigma, \omega)$ in terms of $c$ and $\kappa$ using the Jacobi identities. As in \cite{Gupta:2022mdt} we schematically denote the Jacobi identity involving the modes of the operators $A(z), B(z), C(z)$ as identity $(ABC)$. There are a total of 10 identities to be considered. The four identities $(TTT)$, $(TTH)$, $(THH)$, $(HHH)$ are automatically satisfied. Next there are three with one $P^a$ each: $(TTP^a)$, $(THP^a)$, $(HHP^a)$ -- which can also be seen to be satisfied identically. 

The remaining three Jacobi identities contain at least two $P^a_n$'s: $(TPP)$, $(HPP)$ and $(PPP)$. These involve commutators between two $P^a_m$'s. To impose these we need the commutators of the modes of the simple primaries $(L_m, H_m, P^a_m)$ with $((H^2)_n, \Lambda_n, \Sigma_n)$.
\subsubsection*{Commutators with $(H^2)_n$}
It is easy to see
\bea
[L_m, (H^2)_n] &=& (m-n) \, (H^2)_{m+n} + \frac{\kappa}{6}  \, m(m^2-1) \, \delta_{m+n,0} \, ,\cr
[H_m, (H^2)_n] &=& 2\kappa \, m \, H_{m+n} \, .
\eea
For $[P^a_m, (H^2)_n] $ we first note that the OPE between $P^a(z)$ and $(H^2)(w)$ can be written in terms of quasi-primaries and their descendants as follows:
\bea
P^a(z) (H^2)(w) \sim \frac{P^a(w)}{(z-w)^2} +  \frac{\partial P^a(w)}{2 (z-w)} -2 {\epsilon_b}^a \frac{[P^bH](w)}{z-w}
\eea
where
\bea
[P^aH](z) := (P^aH)(z) + \frac{3}{4} {\epsilon_b}^a \partial P^a
\eea
is a primary. In obtaining this we used the identity: $(HP^a)(z) - (P^aH)(z) -{\epsilon_b}^a \partial P^b(z) = 0$. Using this OPE we can immediately write the commutator of the corresponding modes as
\bea
\label{phsqcomm}
[P^a_m, (H^2)_n] 
&:=& -2{\epsilon_b}^a [P^bH]_{m+n} + \frac{1}{2} (m-n) \, P^a_{m+n} \, .
\eea
\subsubsection*{Commutators with $\Lambda_n$}
It is straightforward to find
\bea
[L_m, \Lambda_n] 
&=& (2m-n) \Lambda_{m+n} + \frac{1}{12}(c+2) m (m^2-1) \, H_{m+n} \, \cr
&& \cr
%
[H_m, \Lambda_n] &=& \kappa \, m \, L_{m+n} + m \, (H^2)_{m+n} \, .
\eea
The next non-trivial commutator is $[P^a_m, \Lambda_n]$. For this again we write the corresponding OPE $P^a(z)\Lambda(w)$ in terms of quasi-primaries and their global descendants. We find
\bea
\label{palambda-ope}
P^a(z) \Lambda(w) &\sim& \frac{-2 {\epsilon_b}^a }{(z-w)^3} \left[P^b(w) + \frac{1}{4}(z-w)  \partial P^b(w) + \frac{1}{20} (z-w)^2 \partial^2 P^b(w) \right] \cr && \cr
&& + \frac{2 [P^aH](w)}{(z-w)^2} + \frac{2 \partial [P^aH](w)}{3 (z-w)} - {\epsilon_b}^a \frac{[TP^b](w)}{z-w} + \frac{ [\partial P^a H](w)}{3(z-w)} \cr &&
\eea
where the relevant composite quasi-primaries are 
\bea
h=3 ~~ &:& ~~  [P^aH](z) := (P^a H)(z)+ \frac{3}{4}  {\epsilon_b}^a \partial P^b(z), \cr
&& ~~~ [HP^a](z) := (HP^a)(z)-\frac{1}{4} {\epsilon_b}^a \partial P^b(z) , \cr
h=4 ~~ &:& 
~~ [H\partial P^a](z) := (H\partial P^a)(z) - 2 \, (P^a\partial H)(z) - \frac{11}{10} \partial^2 P^b(z) {\epsilon_b}^a \cr
&& ~~~ = (\partial P^a H)(z) -2 (P^a \partial H)(z) - \frac{3}{5}  \partial^2 P^b {\epsilon_b}^a := [\partial P^aH](z) , \cr
h=4 ~~ &:&~~ [TP^a](z) := (TP^a)(z) - \frac{3}{10} \partial^2 P^a(z) = (P^aT)(z) - \frac{3}{10} \partial^2 P^a(z) := [P^aT](z) , \cr &&
\eea
%
%
Converting the OPE (\ref{palambda-ope}) into a commutator we find
\bea
[P^a_m, \Lambda_n] &=& - \, {\epsilon_b}^a  [TP^b]_{m+n} +\frac{1}{3} [H\partial P^a]_{m+n} \cr
&& -\frac{1}{10}(6m^2+n^2-3mn-4) {\epsilon_b}^a P^b_{m+n} +\frac{2}{3}(2m-n) [P^aH]_{m+n}   \, .
\eea
\subsection*{Commutators with $\Sigma_n$}
It is easy to see that 
\bea
[L_m, \Sigma_n] 
&=& (2m-n) \Sigma_{m+n} + \frac{1}{4} m (m^2-1) \, H_{m+n} , \cr
[H_m, \Sigma_n] &=& \frac{3}{2} m \, (H^2)_{m+n} \, .
\eea
To find $[P^a_m, \Sigma_n]$ we again write the OPE:
{\small
	\bea
	\label{pasigma-ope}
	P^a(z) \Sigma(w) &\sim& \frac{- {\epsilon_b}^a }{2\kappa (z-w)^3} \left[ P^b(w) + \frac{1}{4} (z-w) \partial P^b(w) + \frac{1}{20} (z-w)^2 \partial^2 P^b(w) \right] \cr
	&& + \frac{3}{2\kappa (z-w)^2}  \left[ [P^aH](w) + \frac{1}{3} (z-w) \partial [P^aH](w) \right] - \frac{1}{4\kappa (z-w)} [\partial P^aH](w) \cr
	&&  - \frac{1}{2\kappa(z-w)} {\epsilon_b}^a \Big[ [HHP^b](w) + [P^aHH](w) + [HP^aH](w) \Big] 
	\eea
}
where the additional quasi-primaries with $h=4$ are:
{\small 
	\bea
	[HHP^a](z) &:=& ((H^2)P^a)(z) -{\epsilon_b}^a \frac{3}{4} (H \partial P^b)(z) -{\epsilon_b}^a \frac{3}{2} ( P^b \partial H)(z) - \frac{27}{40} \partial^2 P^a (z) \, , \cr
	[HP^aH](z) &:=& ((HP^a)H)(z) +\frac{5}{4} {\epsilon_b}^a (H\partial P^b)- \frac{1}{2} {\epsilon_b}^a (P^b \partial H)(z) - \frac{1}{40} (11+20\kappa) \partial^2 P^a , \cr
	[P^aHH](z) &:=& ((P^aH)H)(z) + 4 {\epsilon_b}^a (P^b\partial H)(z) - \frac{1}{10}(5\kappa -17) \partial^2 P^a ,
	\eea
}
and we used the identities
\bea
(HP^a)(z) - (P^aH)(z) - {\epsilon_b}^a \partial P^b (z) &=& 0 \, ,\cr
(\partial P^a H)(z) - (H \partial P^a)(z) +  \frac{1}{2} {\epsilon_b}^a \partial^2 P^b(z) &=& 0 \, .
\eea
Converting the OPE (\ref{pasigma-ope}) into commutator modes we find:
\bea
[P^a_m, \Sigma_n] &=& - \frac{1}{40 \kappa} {\epsilon_b}^a (6m^2+n^2-3mn-4) P^b_{m+n} + \frac{1}{2\kappa} (2m-n) [P^aH]_{m+n} \cr
&& - \frac{1}{4\kappa} [\partial P^a H]_{m+n} - \frac{1}{2\kappa} {\epsilon_b}^a \Big( [HHP^b]_{m+n} + [HP^bH]_{m+n} + [P^b HH]_{m+n} \Big) \cr &&
\eea
We are ready to impose the $(TPP)$ and $(HPP)$ identities. The computations are straightforward and we find that $(TPP)$ identity requires 
\bea
\label{conds11}
2\alpha-c\beta-2\kappa \, \gamma = 0, ~~~ 24 \delta - (2+c)\sigma - 3\omega = 0 \, .
\eea
The $(HPP)$ identity requires
\bea
\label{conds12}
\beta +3 \delta +2\kappa \, \gamma &=& 0, ~~~
\alpha + 6 \kappa \, \delta = 0 \, , \cr
2\beta +\kappa \, \sigma &=& 0, ~~~
2\gamma + \sigma + \frac{3}{2} \omega = 0 \, .
\eea
Finally we need to impose $(PPP)$ Jacobi identity. For this, first we write $[P^c_k, [P^a_m, P^b_n]] $ as
%
{\small
	\bea
	&& \left( \beta + \frac{\gamma}{2} \right) \eta^{ab} (m-n)(k-m-n) P^c_{m+n+k} \cr
	&& - \Big[ \delta \, (m^2+n^2-m n -1) + \frac{1}{10} \left(\sigma + \frac{\omega}{4\kappa} \right) \Big(6k^2+(m+n)^2-3k(m+n)-4 \Big) \Big]  \epsilon^{ab} {\epsilon_d}^c P^d_{m+n+k} \cr
	&& + \left(\frac{\omega}{2\kappa}+ \frac{2\sigma}{3}\right) (2k-m-n) \epsilon^{ab} [P^cH]_{m+n+k} -2\gamma (m-n) \eta^{ab} {\epsilon_d}^c [P^dH]_{m+n+k} \cr
	&& - \sigma \epsilon^{ab} \, {\epsilon_d}^c [TP^d]_{m+n+k}  + \left( \frac{\sigma}{3} - \frac{\omega}{4\kappa} \right)  \epsilon^{ab}  [H\partial P^c ]_{m+n+k} \cr 
	&&-\frac{\omega}{2\kappa} \epsilon^{ab} {\epsilon_d}^c \Big( [HHP^b]_{m+n+k} + [HP^bH]_{m+n+k} + [P^b HH]_{m+n+k} \Big) \, .
	\eea
}
The Jacobi identity to be imposed is
%
	\bea
	[P^c_k, [P^a_m, P^b_n]] &+& \Big((c,k) \rightarrow (a,m) \rightarrow (b,n) \rightarrow (c,k) \Big) \cr
	&+& \Big( (c,k) \rightarrow (b,n) \rightarrow (a,m) \rightarrow (c,k) \Big) =0 .
	\eea
%
The terms that contain $\epsilon^{ab} {\epsilon_d}^c {\cal O}^d_{m+n+k}$ and $\epsilon^{ab} {\cal O}^c_{m+n+k}$ can be dropped as they vanish after summing over the cyclic permutations above. Then there are two types of terms left which contain (i) $P^c_{k+m+n}$, (ii) $[P^c H]_{m+n+k}$. 
\begin{enumerate}
	\item Setting terms containing $[P^aH]_{m+n+k}$ to zero we find one condition 
	\bea
	\label{conds13}
	4\kappa(\gamma-\sigma)-3\omega = 0 .
	\eea
	\item Setting terms containing $P^a_{m+n+k}$ to zero gives the last condition
	\bea
	\label{conds14}
	4\kappa(2\beta+\gamma-2\delta+\sigma)+\omega =0 .
	\eea
\end{enumerate}
\section{Chiral conformal extensions of $\mathfrak{so}(2,2)$}
\label{appendixA}
In this appendix we construct the chiral conformal/${\cal W}$-algebra extensions of $\mathfrak{so}(2,2)$. Repeating the steps of section \ref{so23avatars} we can take the $L_0$ to belong to the Cartan subalgebra and we choose it such that the eigenvalues are real. Taking the generator to be $L_{\mu\nu}$ for $\mu, \nu \in \{0,1,2,3\}$ we take 
\bea
L_0 = \lambda L_{01} + \lambda' \, L_{23}
\eea
We can using $O(2,2)$ transformations 
\begin{enumerate}
\item $(\lambda, \lambda') \leftrightarrow (-\lambda, \lambda')$ using $\Lambda = {\rm diag.}(1,-1,1,1)$
\item $(\lambda, \lambda') \leftrightarrow (-\lambda, \lambda')$ using $\Lambda = {\rm diag.}(1,1,-1,1)$
\item $(\lambda, \lambda') \leftrightarrow (\lambda', \lambda)$ using $\Lambda_{mn} = (-1)^m \delta_{m+n,3}$
\end{enumerate}
So we can again choose $\lambda \ge \lambda' \ge 0$. Then the existence of $L_{\pm 1}$ demands that $(\lambda, \lambda')$ satisfy one of the following four conditions: (i) $\lambda + \lambda' =1$, (ii) $\lambda - \lambda' =1$, (iii) $\lambda + \lambda' =-1$, (iv) $\lambda - \lambda' =-1$. Since we have already assumed $\lambda \ge \lambda' \ge 0$ we need to consider only two cases: (i) $\lambda + \lambda' =1$ and (ii) $\lambda - \lambda' =1$. In each case we can solve the constraints to find the following results:
\begin{enumerate}
\item $\lambda =1 ~~ \& ~~ \lambda' =0$.
\bea
L_0 &=& L_{01}, ~~ (b_{03}^2-b_{12}^2) L_1 = b_{12} (L_{02}-L_{12}) + b_{03} (L_{03}-L_{13}) \cr
L_{-1} &=& b_{12} (L_{02}+L_{12}) + b_{03}(L_{03}+L_{13})
\eea
The residual $O(2,3)$ transformations (that keep these form-invariant) can be seen to leave $(b_{03}^2-b_{12}^2)$ invariant up to a positive scaling : $(b_{03}^2-b_{12}^2) \rightarrow e^\beta (b_{03}^2-b_{12}^2)$. Using the discrete elements of $O(2,3)$ we can change the signs of $b_{03}$ and $b_{12}$ independently. Thus using these residual transformations we can choose $(b_{03}, b_{12})$ to be either $(1,0)$ or $(0,1)$.
\item $\lambda =1/2= \lambda' $
\bea
L_0 &=& \frac{1}{2} (L_{01}+L_{23}), ~~ L_1 = -a_{12} (L{02}+L_{03}-L_{12}-L_{13}), \cr
L_{-1} &=& \frac{1}{4 a_{12}} (L_{02}-L_{03}+L_{12}-L_{13})
\eea
The residual transformations can be used to fix $a_{12} = 1/2$. 
\end{enumerate}
This analysis gives us three inequivalent embeddings of $\mathfrak{sl}(2, {\mathbb R})$ in $\mathfrak{so}(2,2)$ -- resulting in three different avatars of $\mathfrak{so}(2,2)$.
\begin{enumerate}
\item $\mathfrak{h} = \emptyset$ with $(b_{03}, b_{12}) = (1,0)$
\bea
L_1 &=& L_{03}-L_{13}, ~~ L_0 = L_{01}, ~~ L_{-1} = L_{03}+L_{13}, \cr
P_1 &=& L_{02}-L_{12}, ~~ P_0 = L_{23}, ~~ P_{-1} = -L_{02}-L_{12}.
\eea
{\small
\bea
[L_m, L_n]&=& (m-n) \, L_{m+n} = [P_m, P_n] , ~~ [L_m, P_n] = (m-n) P_{m+n},  \cr
&&
\eea
}
\item $\mathfrak{h} = \emptyset$ with $(b_{03}, b_{12}) = (0,1)$
\bea
L_1 &=& -L_{02}+L_{12}, ~~ L_0 = L_{01}, ~~ L_{-1} = L_{02}+L_{12}, \cr
P_1 &=& L_{03}- L_{13}, ~~ P_0 = -L_{23}, ~~ P_{-1} = L_{03}+L_{13}.
\eea
{\small
\bea
[L_m, L_n]&=& (m-n) \, L_{m+n} = [P_m, P_n] , ~~ [L_m, P_n] = (m-n) P_{m+n},  \cr
&&
\eea
}
\item $\mathfrak{h} = \mathfrak{sl}(2, {\mathbb R})$
\bea
L_1 &=& -\frac{1}{2} (L_{02}+L_{03}-L_{12}-L_{13}), \cr
L_{-1} &=& \frac{1}{2} (L_{02}-L_{03}+L_{12}- L_{13}), ~~ L_0 = \frac{1}{2} (L_{01}+L_{23}) \cr
J_1 &=& \frac{1}{2} (-L_{02}+L_{03}+L_{12}-L_{13}), \cr 
J_0 &=& \frac{1}{2} (L_{01}-L_{03}), ~~ L_{-1} = \frac{1}{2} (L_{02}+L_{03}+L_{12}+L_{13}).
\eea
\bea
[L_m, L_n] = (m-n) L_{m+n}, ~~ [J_a, J_b] = (a-b) J_{a+b}, ~~ [L_m, J_a]  =0
\eea
\end{enumerate}
Unlike in the $\mathfrak{so}(2,3)$ case the first two avatars above with $\mathfrak{h} = \emptyset$, even though cannot be mapped to each other by any $O(2,3)$ transformations nevertheless give rise to identical algebras. So we have two distinct avatars of $\mathfrak{so}(2,2)$ which can be extended to the following chiral algebras:
\begin{enumerate}

\item $\mathfrak{h} = \mathfrak{sl}(2, {\mathbb R})$
\bea
[L_m, L_n] &=& (m-n) L_{m+n} + \frac{c}{12} m (m^2-1) \delta_{m+n,0}, \cr
[J_{a,m}, J_{b,n}] &=& (a-b) J_{a+b,m+n} - \frac{1}{2}\kappa (a^2+b^2-ab-1) m \, \delta_{m+n,0}, \cr
[L_m, J_{a,n}] &=& -n \, J_{a,m+n}
\eea
The Jacobi identity imposes no constraints on $c$ and $\kappa$.
\item $\mathfrak{h} = \emptyset$
\bea
[L_m, L_n] &=& (m-n) L_{m+n} + \frac{c}{12} m (m^2-1) \delta_{m+n,0}, \cr
[L_m, P_n] &=& (m-n) P_{m+n} + \frac{\tilde c}{12} m (m^2-1) \delta_{m+n,0}, \cr
[P_m, P_n] &=&  (m-n) L_{m+n}  + \gamma \, m(m^2-1) \delta_{m+n,0}
\eea
Imposing the Jacobi identity here leads to the condition that $\gamma = \frac{c}{12}$. 
\end{enumerate}
Thus we arrive at the result that there are exactly two chiral conformal algebra extensions of $\mathfrak{so}(2,2)$. The first one is the CIG algebra \cite{Avery:2013dja, Apolo:2014tua} (however, with no relation between $c$ and $\kappa$) and the second can be mapped to the non-chiral Brown-Henneaux one (with no relation between left and right Virasoro central charges) after appropriate redefinitions. Specifically, if we take $L_n \rightarrow \ell_n+ \bar \ell_n$ and $P_n \rightarrow \ell_n-\bar \ell_n$ we will recover two commuting copies of Virasoros generated by $\ell_n$ with central charge $(c+\tilde c)/2$ and $\bar \ell_n$ with central charge $(c-\tilde c)/2$ respectively. It is interesting that even though we started with chiral conformal/{\cal W}-algebra extensions we seemed to get both chiral and non-chiral extensions of $\mathfrak{so}(2,2)$ already known. 

\bibliographystyle{utphys}
\providecommand{\href}[2]{#2}\begingroup\raggedright\endgroup 
\end{document}